\documentclass{aa}

\usepackage{graphicx}

\begin{document}

\title{Swirling astrophysical flows --- efficient amplifiers of Alfv\'en waves!?}

\author{Andria D. Rogava\inst{1,2,3}\thanks{On leave from Abastumani
                  Astrophysical Observatory, Kazbegi ave.~2a,
                  Tbilisi--380060, Georgia}
   \and Swadesh M. Mahajan\inst{4}
   \and Gianluigi Bodo\inst{2}
   \and Silvano Massaglia\inst{1}}

\institute{Dipartimento di Fisica Generale,
           Universit\'a degli Studi di Torino, Via Pietro Giuria 1, Torino
           I-10125, Italy
 \and Osservatorio Astronomico di Torino, Strada
           dell'Osservatorio 20, I-10025, Pino Torinese, Italy
 \and Abdus Salam International Centre for Theoretical Physics,
           Strada Costiera 11, Trieste I-34014, Italy
 \and Institute for Fusion Studies, The University of Texas at Austin,
           Texas 78712, U.S.A.
    }

\offprints{Andria Rogava}

\date{Received / Accepted }

\authorrunning{Rogava et al.}

\titlerunning{Swirling astrophysical flows...}

\abstract{We show that a {\it helical} shear flow of a magnetized
plasma may serve as an efficient amplifier of Alfv\'en waves. We
find that even when the flow is purely ejectional (i.e., when no
rotation is present) Alfv\'en waves are amplified through the
transient, shear-induced, algebraic amplification process. Series
of transient amplifications, taking place sequentially along the
flow, may result in a {\it cascade amplification} of these waves.
However, when a flow is swirling or {\it helical} (i.e., some
rotation is imposed on the plasma motion), Alfv\'en waves become
subject to new, much more powerful shear instabilities. In this
case, depending on the type of differential rotation, both usual
and parametric instabilities may appear. We claim that these
phenomena may lead to the generation of large amplitude Alfv\'en
waves and the mechanism may account for the appearance of such
waves in the solar atmosphere, in accretion-ejection flows and in
accretion columns. These processes may also serve as an important
initial (linear and nonmodal) phase in the ultimate subcritical
transition to MHD Alfv\'enic turbulence in various kinds of
astrophysical shear flows.  }

\maketitle

\section{Introduction}


Most of astrophysical objects involve different kinds of plasma
flows. Recently it was fully realized that collective phenomena in
flows with spatially inhomogeneous velocities ({\it shear flows}
hereafter referred as SF) are characterized by remarkable, so
called ``nonmodal" processes, related with non-self-adjointness of
linear dynamics of perturbations. Namely, it was found that SF:
exchange energy with sound waves (\cite{bf92}); couple different
collective modes with one another and lead to their mutual
transformations (\cite{crt96}); generate nonperiodic, vortical
modes of collective behaviour (so called ``shear vortices"
\cite{rcm98}) - which eventually may or may not acquire wave-like
features; excite {\it beat wave} phenomena both in neutral fluids
(\cite{rm97}) and plasmas (\cite{prm98}).

These processes take place not only in neutral fluids (\cite{bf92,
rm97}), standard MHD (\cite{crt96, rmb96, tvy01}) and
electrostatic plasmas (\cite{rcm98, vyt00, mms00}), but also in
strongly magnetized plasmas with anisotropic pressure
(\cite{crt97}), electron-positron plasmas (\cite{mmr97}) and dusty
plasmas (\cite{pkr00}). The possible role of these phenomena in
astrophysical context was immediately realized and a number of
astrophysical applications, including pulsar magnetospheric
plasmas (\cite{mmr97}), solar atmospheric phenomena
(\cite{prm98,rpm00}) and galactic gaseous disk dynamics
(\cite{rph99}) appeared within a span of years.

One of the main shortcomings of all these studies, stemming from
the very design of nonmodal schemes, is that the description  is
given in the wave number space (${\bf k}-space$),  the knowledge
about the  appearance of shear-induced phenomena in the real,
physical space is lacking. Two more serious limitations of these
investigations are related with the neglect of the back reaction
of perturbations on the mean flow and with the omission of viscous
dissipation effects. However recently a successful effort was made
in the direction of spatial visualization of the shear-induced
wave transformations (\cite{bprr01}), became clear that
shear-induced processes in SF are quite robust and easily
recognizable, even in the presence of quite heavy dissipation.

All these processes were  studied predominantly for simple,
plane-parallel flow geometries with linear velocity shear
profiles. However, recently, a new method was developed
(\cite{mr99}), which allows a {\it local} analysis of the dynamics
of linearized perturbations in SF with arbitrarily complex
geometry and kinematics ${\bf U}(x,y,z)$. It was found that a
slight deviation from the plane-parallel mode of motion brings
into the game a variety of new, exotic and asymptotically
persistent modes of collective behavior.  Still the systematic
investigation of shear-induced processes in kinematically complex
SF is a challenging task in a state of infancy. The original study
has been limited to two-dimensional flow patterns of neutral
fluids (\cite{mr99}).

One begins to wonder whether those astrophysical systems, where
kinematically complex modes of plasma motion are definitely
present --- e.g., astrophysical jets (\cite{f98}) or solar
tornados (\cite{pm98, vl99}) --- might sustain these collective
phenomena and what kind of observational consequences could they
lead to.

In this paper we consider interaction of a helical flow with
Alfv\'en waves generated within the flow. This problem is actual
in plasma physics and plasma astrophysics from
experimental/observational, theoretical and numerical points of
view.

We are used to observe Alfv\'en waves, being in interaction with
plasma flows, in different astrophysical situations. For example,
for the long time it is known  that the Sun radiates Alfv\'en
waves: outward propagating Alfv\'en waves are routinely observed
in the solar wind flow at $r>0.3\,$AU (\cite{h90}). The
observations of quasiperiodic pulsed ionospheric flows (PIF) have
shown that the PIF are driven and correlated with Alfv\'enic
fluctuations observed in the upstream solar wind (\cite{ppmy02}).
While the observations of the solar transition region (\cite{p01})
imply that structuring of the transition region involves closed
loops and coronal funnels showing unambiguous evidence for the
presence of passing Alfv\'en waves.

In the theoretical domain the problem of the interaction between
Alfv\'en waves and ambient flows is quite popular topic of studies
in the wide range of applications including different kinds of
laboratory, geophysical and astrophysical plasma flows. Recent
studies of toroidal flows in axisymmetric tokamaks (\cite{vbb00}),
for instance, revealed that these flows generate low-frequency
Alfv\'en waves. The problem of Alfv\'en waves sustained by plasma
flows is especially popular in the context of solar physics. It is
argued that chromosphere and transition region flows are primary
energy sources for the fast solar wind; while Alfv\'en waves,
generated by post-reconnection processes, are continuously
interacting with these flows (\cite{rhwt01}). In accretion disks
Alfv\'en waves also seem to be actively interacting with the disk
flow - they are found to be unstable both in high-$\beta$
(\cite{bh91}) and low-$\beta$ (\cite{tpc92}) disks. Most recently
it was claimed that the Accretion-Ejection Instability
(\cite{tp99}) can extract accretion energy and angular momentum
from magnetized disk, generate Alfv\'en waves and ``feed" the disk
corona with them (\cite{vt02}).

Currently sophisticated numerical codes are developed, which allow
to study the propagation of Alfv\'en waves along an open magnetic
flux tube (\cite{sks01}). These simulations are aimed to clarify
mechanisms of the coronal heating and of the formation of solar
plasma flow patterns: viz. solar spicules, macrospicules and solar
tornadoes. More general numerical tools, like FINESSE and PHOENIX,
aimed to study waves and instabilities in different kinds of
flows, were very recently developed (\cite{bbgvk02}). We assume
that all these numerical tools could also be used in a number of
astrophysical applications involving ``parent" flows and ``inborn"
Alfv\'en waves interacting with each other.

Most of above-cited studies considered simple (plane-parallel or
locally plane-parallel) kinds of flows and the processes were
treated by means of usual normal-mode analysis. The purpose of
this paper (paper I) is to study `nonmodal' evolution of Alfv\'en
waves in swirling flows and to show that these flows may operate
as efficient ``amplifiers" of Alfv\'en waves.

In the next section we develop the mathematical formalism and
derive general equations governing the evolution of Alfv\'en waves
in helical flows. However, the third section is dedicated to the
study of the simpler example of a parallel SF (`pure outflow").
Still, even in this simple case, we find that the amplification of
Alfv\'en waves takes place. The amplification mechanism is linear,
transient and can be described by appealingly simple mathematics.
The amplification occurs within a relatively brief time interval
and it appears as an abrupt, burst-like increase of the wave
amplitude. It is tempting to argue that individual acts of wave
amplifications, occurring sequentially, may ignite the ``chain
reaction" of {\it nonmodal cascade} amplification of Alfv\'en
waves.

The fourth section of the paper is dedicated to fully helical SF.
We show that when SF are helical (``outflow + rotation") there
appear new classes of shear instabilities capable of generating
high-amplitude Alfv\'en waves. The presence of differential
rotation is crucial for these instabilities and in certain cases
the instability is of parametric nature. These instabilities are
powerful, often  have a `resonant' nature and are exclusively
related with the kinematic complexity of `parent' shear flows.

The fifth section of the paper contains discussion of the obtained
results and of their possible astrophysical applications. We argue
that such wave amplification processes may have various
astrophysical manifestations including large amplitude Alfv\'en
waves actually observed in the solar atmosphere (\cite{bst95}).
They could provide the necessary seeding for the development of
MHD turbulence in hydromagnetic shear flows. These processes may
be present and may lead to perceptible morphological variety and
diverse observational appearances in various kinds of {\it
`accretion-ejection'} flows: innermost regions of accretion disks,
disk-jet transition regions, accretion columns in X-ray pulsars
and cataclysmic variables, inner regions of galactic gaseous
disks, etc.

This paper will be followed by the second one (paper II), where we
will consider the same flow structure but allow the perturbations
to be fully compressible, bringing on-stage another two additional
linear MHD wave modes - the slow and fast magnetosonic waves.

\section{Main consideration}

Our aim is to develop a theory of collective phenomena in helical
flows of magnetized plasmas. We adopt the standard MHD model and
consider {\it incompressible} perturbations in  an axisymmetric,
cylindrical, steady flow of a plasma with uniform density
(${\rho}_0=const$), embedded in a vertical, homogeneous ${\bf
B}_0=(0,~0,~B_0)$ magnetic field.  With $\rho = {\rho}_0+
\varrho$, $P=P_0+p$, ${\bf V}={\bf U}+{\bf u}$, ${\bf B}={\bf
B}_0+ {\bf B}^{'}$, the basic set of MHD equations for linearized
perturbations is:
$$ \nabla \bf u=0, \eqno(1)
$$
$$
D_t \bf u+({\bf u} \cdot {\nabla}){\bf U}=-{1 \over
{\rho}_0}{\nabla}p-{{{\bf B}_0}\over
{4{\pi}{\rho}_0}}{\times}({\nabla}{\times}{\bf B}^{'}), \eqno(2)
$$
$$
D_t{\bf B}^{'}=({\bf B}^{'} \cdot {\nabla}){\bf U} + ({\bf
B}_0 \cdot {\nabla}){\bf u}, \eqno(3)
$$
$$
\nabla {\bf B}^{'}=0, \eqno(4)
$$
where $D_t{\equiv}{\partial}_t+({\bf U} \cdot
{\nabla})$ is the convective derivative operator and ${\bf U}(r)$
is the vector field of the {\it steady state flow}. Unfortunately
the actual kinematic portrait of astrophysical helical flows is
largely unknown. There is no credible data for recently
discovered solar macrospicules with helical plasma motion ({\it
solar tornados}), for {\it accretion columns} related to X-ray
pulsars and cataclysmic variables, for {\it accretion-ejection flows} and {\it galactic and
extragalactic jets}. Therefore, we need to adopt a
phenomenological model for ${\bf U}$, that has to be general
enough to encompass different possible sorts of real helical
flows. Preferably the model must be three-dimensional and it
should include a possibility of both outflowing/inflowing and
rotational modes of motion.  The model we choose is (\cite{rp99}):



$$
{\bf U}(r){\equiv}[0,~r{\Omega}(r),~U(r)], \eqno(5)
$$
where $r=(x^2+y^2)^{1/2}$ is a distance from the rotation axis.
For the angular velocity we take:
$$
{\Omega}(r)={\cal A}/r^n, \eqno(6)
$$
with ${\cal A}$ and $n$ as constants.  This model implies the
following two limiting cases:

\begin{itemize}

\item
{\it Rigid rotation}, with $n=0$ and ${\cal A}={\Omega}_0$.

\item
{\it Keplerian rotation}, with $n=3/2$ and ${\cal A}=(GM)^{1/2}$.

\end{itemize}

The corresponding Cartesian components of the linear azimuthal
velocity $U_{\phi}=r{\Omega}(r)$ are:
$$
U_x(x,y)=-{\cal A}y/r^n, \eqno(7a)
$$
$$
U_y(x,y)={\cal A}x/r^n.  \eqno(7b)
$$

A basic role in our analysis is played by the shear matrix ${\cal S}_{ik}
{\equiv}{\partial}U_i/{\partial}x_k$.
Evidently rotational part of the flow velocity generates the following four nonzero
components of the shear
matrix:
$$
{\sigma}{\equiv}{\cal S}_{xx}=2{\cal A}nx_0y_0/r_0^{n+2}, \eqno(8a)
$$
$$
{\cal S}_{yy}=-{\sigma}, \eqno(8b)
$$
$$
A_1{\equiv}{\cal S}_{xy}=-{\cal A}[x_0^2+(1-n)y_0^2]/r_0^{n+2}, \eqno(8c)
$$
$$
A_2{\equiv}{\cal S}_{yx}={\cal A}[(1-n)x_0^2+y_0^2]/r_0^{n+2}, \eqno(8d)
$$

As regards the outflow component of the velocity we can use the model (\cite{rpm00}):
$$
U_z(x,y)=U_m[1-(r/R)^2], \eqno(9a)
$$
comprising a parabolically
sheared ``outflow" along the $Z$ axis, being similar to the well-known Hagen-Poiseuille flow.
The outflow model may be modeled also in a number of other ways. In astrophysical jet
literature, for instance, the model (\cite{f98}):
$$
U_z=U_m/cosh[r^m], \eqno(9b)
$$
is often used. One can easily see
that the outflow component, given by (9), generates the following
two components of the shear matrix:

$$
C_1{\equiv}{\cal S}_{zx}=-2U_mx_0/R^2, \eqno(10a)
$$
$$
C_2{\equiv}{\cal S}_{zy}=-2U_my_0/R^2, \eqno(10b)
$$

Therefore, the resulting helical flow is the superposition of a
cylindrical outflow along the $Z$ axis
with a differential motion in the
transverse cross-section of the jet. The complete $3 \times 3$
traceless shear matrix is:
$$
{\cal S}={\left(\matrix{{\sigma}&A_1&0\cr A_2&-{\sigma}&0 \cr
C_1&C_2&0\cr}\right)}. \eqno(11)
$$

This shear matrix has a number of interesting properties. Its square is
equal to:
$$
||{\cal S}^2||={\left(\matrix{
                               {\Gamma}^2     &           0     & 0 \cr
                               0              & -{\Gamma}^2     & 0 \cr
                               {\varepsilon}_1& {\varepsilon}_2 &0  \cr}
\right)}, \eqno(12)
$$
where we use the notation:
$$
{\varepsilon}_1{\equiv}A_2C_2+{\sigma}C_1, \eqno(13a)
$$
$$
{\varepsilon}_2{\equiv}A_1C_1-{\sigma}C_2; \eqno(13b)
$$
$$
{\Gamma}{\equiv}({\sigma}^2+A_1A_2)^{1/2}=\sqrt{n-1}~{\Omega}(r),
\eqno(13c)
$$
and its cube is: $||{\cal S}^3||={\Gamma}^2||{\cal S}||$.

The nonmodal local method for studying the dynamics of linearized
perturbations in kinematically complex flows (\cite{l84}, \cite{cc86},
\cite{mr99}) allows to reduce the initial set of partial differential equations
for perturbation variables $F({\bf r},t)$, defined in the real physical space,
to the initial value
problem formulated for the perturbation variable amplitudes
${\hat F}({\bf k},t)$, defined in the space
of wave numbers ({\bf k}-space). The key element of this approach is the
time variability of ${\bf k}$'s, imposed by the presence of the shear flow!
This variability is governed by the following set of equations:
$$
{\bf k}^{(1)}+{\cal S}^T\cdot{\bf k}=0,  \eqno(14)
$$
henceforth for an arbitrary function $f$ we use the notation: ${\partial}_t^nf{\equiv}f^{(n)}$.

For the helical flow we find that ${\bf k}^{(3)}={\Gamma}^2{\bf k}^{(1)}$.
The corresponding characteristic equation $det[{\cal S}^T-{\lambda}{\times}
{\cal I}]=0$ (with ${\cal I}$ being a unit matrix) yields the equation for the
eigenvalues:
$$
{\lambda}({\lambda}-{\Gamma})({\lambda}+{\Gamma})=0, \eqno(15a)
$$
with the solutions:
$$
{\lambda}_1=0,~~~~{\lambda}_{2,3}={\pm}{\Gamma}. \eqno(15b)
$$

We see that the differential rotation parameter $n$
plays a decisive role in determining the evolution scenario for the
wave number vector ${\bf k}(t)$: when $n<1$ (including the
rigid rotation case) $\Gamma$ is imaginary and the time evolution
of ${\bf k}(t)$ is periodic, while when $n>1$ (including the
Keplerian rotation regime), $\Gamma$ is real and makes the time
behavior of ${\bf k}(t)$ exponential.



In the dimensionless notation $v_{x,y,z}{\equiv}{\hat
u}_{x,y,z}$, ${\cal P}{\equiv}i{\hat p}/{\rho}_0$,
$b_{x,y,z}{\equiv}i{\hat B}^{'}_{x,y,z}/B_0$, perturbation amplitudes
evolve as:
$$
({\bf k}{\cdot}{\bf v})=0, \eqno(16)
$$
$$
{\bf v}^{(1)}+{\cal
S}{\cdot}{\bf v}=-{\bf k}{\cal P}-C_{\rm A}^2[{\bf k}b_z-k_z {\bf b}],
\eqno(17)
$$
$$
{\bf b}^{(1)}={\cal S}{\cdot}{\bf b}-k_z{\bf v},
(i=x,y) \eqno(18)
$$
$$
({\bf k}{\cdot}{\bf b})=0. \eqno(19)
$$

In the next two sections we separately consider  the problem of  a purely
ejectional/injectional flow, and a flow with helical motion
of plasma particles. We shall see that in both cases the Alfv\'en waves sustained by the flow
are amplified. For the former case the amplification is transient (algebraic instability),
while for the latter case the velocity shear makes Alfv\'en waves exponentially unstable.

\section{Ejectional flow}

This case corresponds to $\sigma=A_1=A_2=0$. We can introduce a
vector ${\bf C}{\equiv}[C_x \equiv C_1,C_y \equiv C_2,C_z=0]$,
which is proportional to the position vector ${\bf r}_0= (x_0,
y_0, 0)$ connecting the point $A(x_0,y_0,z_0)$ (the centre of our
local frame) with the $Z$ axis in a $z=z_0$ plane. Its absolute
value is equal to $dU_z/dr$. The wave number dynamics is given by
simple linear solutions: $k_i(t)=k_i(0)-C_ik_zt$, with $i=x,~y$,
while the vertical component $k_z$ stays constant.  The symmetry
of the flow implies that the vector product of  ${\bf C}$ and
${\bf k}$ has a constant vertical component ${\Delta}{\equiv}-
({\bf C}{\times}{\bf k})_z=C_yk_x(t)-C_xk_y(t)=const(t)$, which,
in its turn, is a linear combination of transverse components of
the wave number vector. This fact suggests the decomposition
$k_{\parallel}(t){\equiv}({\bf C} \cdot {\bf k})/|{\bf C}|$,
$k_{\perp}{\equiv}({\bf C} \times
{\bf k})_z/|{\bf C}|$, i.e.,
$$
{\bf k}(t)=(k_{\parallel}(t),~k_{\perp},~k_z), \eqno(20)
$$
with the entire time dependence of ${\bf k}(t)$ contained in
$$
k_{\parallel}(t)=k_{\parallel}(0)-|{\bf C}|k_zt. \eqno(21)
$$


The system admits a surprisingly efficient and complete analytic
description if we introduce, first, vectors of hydrodynamic
(${\bf{\Omega}_h}{\equiv}{\bf k}{\times}{\bf v}$) and magnetic
(${\bf{\Omega}_m}{\equiv}{\bf k}{\times}{\bf b}$) vorticity and
consider the following scalar products: $V{\equiv}({\bf
C}{\cdot}{\bf v}) $, ${\cal V}{\equiv}({\bf
C}{\cdot}{\bf{\Omega}}_h)$, $B{\equiv}({\bf C} {\cdot}{\bf b})$,
and ${\cal B}{\equiv}({\bf C}{\cdot}{\bf{\Omega}}_m)$. The
usefulness of these variables  becomes apparent when we check that
they obey the following set of first-order equations:
$$
V^{(1)}=-[ln(|{\bf k}(t)|^2)]^{(1)}V+C_{\rm A}^2k_zB, \eqno(22a)
$$
$$
{\cal V}^{(1)}={\Delta}V+C_{\rm A}^2k_z{\cal B}, \eqno(22b)
$$
$$
B^{(1)}=-k_zV, \eqno(23a)
$$
$$
{\cal B}^{(1)}=-{\Delta}B-k_z{\cal V}, \eqno(23b)
$$
which is remarkably simple with only {\it one}
time-dependent coefficient (time dependence exclusively due to
shear) in Eq. (22a).

These variables give complete description of the system, because
``physical"
variables ${\bf v}$ and ${\bf b}$ are readily expressed by means of $V$,
$\cal V$,
$B$, and $\cal B$ via the following vector identities:
$$
{\bf v}={1 \over{|{\bf C}{\times}{\bf k}|^2}}{\biggl[{\cal V}({\bf C}{\times}
{\bf k})+V({\bf k}{\times}({\bf C}{\times}{\bf k}))\biggr]},
\eqno(24a)
$$
$$
{\bf b}={1 \over{|{\bf C}{\times}{\bf k}|^2}}{\biggl[{\cal B}({\bf C}{\times}
{\bf k})+B({\bf k}{\times}({\bf C}{\times}{\bf k}))\biggr]}.
\eqno(24b)
$$

While for the kinetic and magnetic energies of the system we also
have rather transparent expressions:
$$
E_{\rm k}={1 \over 2}|{\bf
v}|^2={1 \over{2|{\bf C}{\times}{\bf k}|^2}} {\biggl[{\cal
V}^2+V^2|{\bf k}|^2\biggr]}, \eqno(25a)
$$
$$
E_{\rm m}={C_{\rm A}^2 \over
2}|{\bf b}|^2={C_{\rm A}^2 \over{2|{\bf C}{\times} {\bf
k}|^2}}{\biggl[{\cal B}^2+B^2|{\bf k}|^2\biggr]}, \eqno(25b)
$$
$$
E_{\rm tot}{\equiv}E_{\rm k} + E_{\rm m}.  \eqno(25c)
$$

From (22-23) we can further notice that the Alfv\'en waves, sustained by this
system are ``splitted" by the presence of the velocity shear.

\begin{itemize}

\item

The {\it first} branch comprises just a simple,
constant amplitude and constant frequency Alfv\'en mode, which is
governed by the second order equation for ${\cal B}$
$$
{\cal B}^{(2)}+{\omega}_{\rm A}^2{\cal B}=0, \eqno(26)
$$
with ${\omega}_{\rm A}{\equiv}C_{\rm A}k_z$ and with obvious solution
${\cal B}(t)={\cal C}sin({\omega}_{\rm A}t+{\phi}_0)$.

\item

The second branch\footnote{This branch normally (i.e. in the absence of
shear) also possesses
constant Alfv\'en frequency, but in a certain sense it is the
incompressible limit of the
fast magnetosonic wave.} is heavily modified by the presence of the
velocity shear. Its
temporal evolution is described by the second order equation for $B$:
$$
B^{(2)}+[ln(|{\bf k}(t)|^2)]^{(1)}B^{(1)}+{\omega}_{\rm A}^2B=0. \eqno(27)
$$

We can further simplify this equation in terms of the variable
${\Psi}{\equiv}|{\bf k}|B$ which obeys
$$
{\Psi}^{(2)}+{\biggl[{\omega}_{\rm A}^2-{{k_z^2|{\bf C}{\times}{\bf
k}|^2} \over{|{\bf k}|^4}}\biggr]}{\Psi}=0. \eqno(28)
$$

Introducing the dimensionless parameter
$$
a{\equiv}{{C_{\rm A}}\over{|{\bf
C}|}}\left[k_{\perp}^2+ k_z^2\right] ^{1/2}, \eqno(29)
$$
and the dimensionless ``time" variable:
$$
T{\equiv}-(C_{\rm A}/|{\bf
C}|)k_{||}(t)=C_{\rm A}[k_zt-k_{\parallel}(0)/|{\bf C}|], \eqno(30)
$$
we may transform (28) into the remarkably simple differential
equation:
$$
{{d^2{\Psi}}\over{dT^2}}+{\biggl[1-{{a^2}\over{{\left(a^2+T^2
\right)}^2}} \biggr]}{\Psi}=0. \eqno(31)
$$

Solutions of this differential equation ${\Psi}_a(T)$ give the
complete solution of the system, because all physical variables
are readily expressed through this function, its first
derivative and the trivial solution for the first, non-modified
branch given by ${\cal B}(t)$. It means that the full solution of
the initial value problem is expressible by certain combinations of
these two Alfv\'en modes: one that does not ``feel" the velocity
shear and the another one that is affected by it.

\end{itemize}


Note that the symmetry of (31) implies that its solutions must be
invariant with respect to the inversion operation. Namely, if a
pair of initial values ${\Psi}_a(-T_0)$ and ${\Psi}^{(1)}_a(-T_0)$
gives a certain solution then ${\Psi}_a(T_0)$ and
$-{\Psi}^{(1)}_a(T_0)$ gives inversion-symmetric solution.

\begin{figure}
  \resizebox{\hsize}{!}{\includegraphics[angle=90]{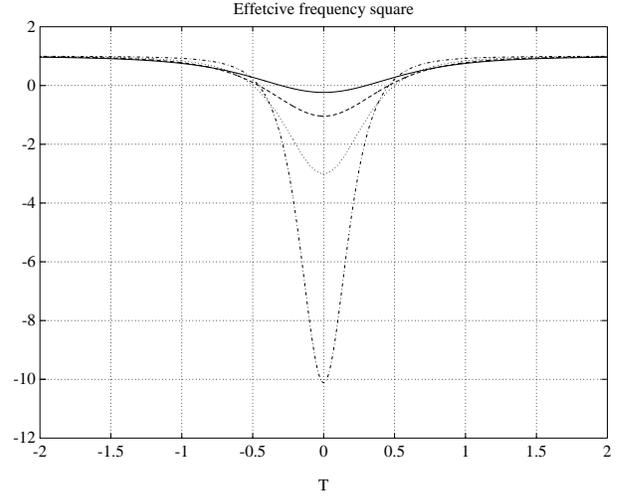}}
  \caption{The plots of the function  ${\Omega}^2_{\rm eff}(T)$ for
different values of the $a$ parameter. Solid line corresponds to
the case $a=0.9$, dashed line to $a=0.7$, dotted one to $a=0.5$,
and dashed-dotted to $a=0.3$.}\label{fig1}
\end{figure}

Analysis of (31) begins by studying the temporal behavior of the
``effective frequency"
$$
{\Omega}_{\rm eff}(T){\equiv}{\sqrt {
1-{{a^2}\over{{\left(a^2+T^2 \right)}^2}}}}. \eqno(32)
$$

For sufficiently large absolute values of $T$,
${\Omega}_{eff}(T)\simeq 1$, implying that  at the corresponding
periods of time, the system supports just the usual constant
frequency and constant amplitude Alfv\'en waves. However, when
$|a|<1$ there are two moments of time $T_{\pm}{\equiv}\pm {\sqrt
{a(1-a)}}$ when the ``effective frequency of shear-modified
Alfv\'en waves" becomes zero and in the interval $T_{-}<T< T_{+}$
it stays imaginary. Therefore,  within this time interval (with
the width determined by $\Delta T{\equiv}2T_{+}$), swift and
sudden changes in the temporal evolution of Alfv\'en waves may be
expected.

Quantitative picture of this behavior is illustrated by Fig.1
where we plotted the  ${\Omega}^2_{\rm eff}(T)$  for different
(but all $|a|<1$) values of $a$.  One sees that with the decrease
of $a$ the minimum of ${\Omega}^2_{\rm eff}(T)$  becomes sharper:
the depth of the minimum in Fig.1 (which accounts for maximum
increments of the transient instability) steadily increases with
the decreasing $a$. The width of the time interval, in which this
function stays negative (and, correspondingly, ${\Omega}_{\rm
eff}(T)$ is imaginary) is maximum for $a=1/2$ and tends to zero
when $a{\simeq}1$ and $a \ll 1$.

At a first glance, from (31), it seems that decreasing $a$ we
would have the continuous increase of the transient amplification
rate. However decreasing $a$ we also make smaller the scaling
factor between the physical time $t$ and the variable $T$, which
appears in (31). Besides, the small values of $a$ imply smaller
values of the Alfv\'en speed and the vertical component of the
wavenumber vector $k_z$. Therefore, we can conclude that transient
amplification factor for lower frequency Alfv\'en waves is higher,
but for the amplification to occur the system needs a longer time
interval. This means, in turn, that {\it outflows with
shorter/longer lifetime values are expected to amplify
higher/lower frequency Alfv\'en waves}.

\begin{figure}
  \resizebox{\hsize}{!}{\includegraphics[angle=90]{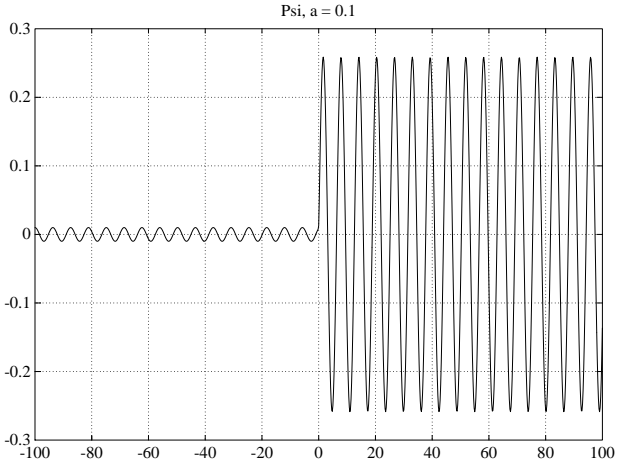}}
  \caption{The numerical solution of Eq.(31) featuring the function
${\Psi}_a(T)$ for $a=0.1$ case. The initial values are:
${\Psi}_a(-100)=0.01$ and ${\Psi}_a^{(1)}(-100)=0$.}\label{fig2}
\end{figure}

\begin{figure}
  \resizebox{\hsize}{!}{\includegraphics[angle=90]{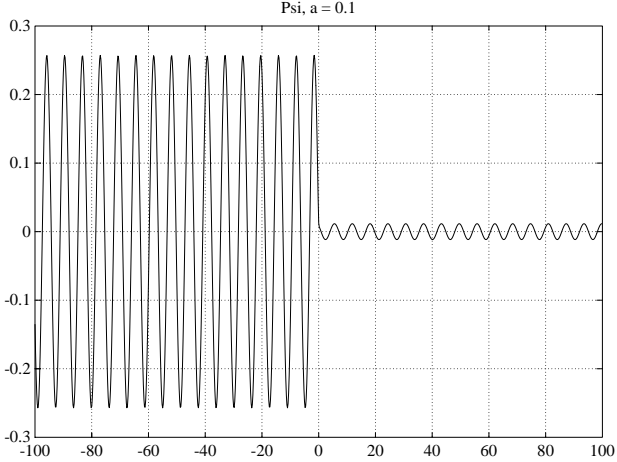}}
  \caption{The numerical solution of Eq.(31), which is
inversion-symmetric to the one given on Fig.2. Initial values here
are: ${\hat{\Psi}}_a(-100)=-0.1354$ and
${\hat{\Psi}}_a^{(1)}(-100)=-0.2186$.}\label{fig3}
\end{figure}

The presence of the imaginary effective frequency for a limited
time interval means that Alfv\'en waves excited and maintained by
the shear flow  become subject to a certain, velocity shear
induced instability, which is present only within the limited time
interval. That is why we specify this phenomenon by the term {\it
``transient instability."} Since the time interval is rather brief
one can expect that the appearance of this instability will have a
burst-like, explosive nature: initially at times $T < T_{-}$ waves
stay almost unaffected by the presence of the shear flow. However,
as soon as the system will enter the transient instability
interval $T_{-}<T< T_{+}$, the Alfv\'en waves undergo drastic and
abrupt change in their amplitudes. Depending on the initial
values, the mode  of evolution will be a certain mixture of
swiftly decaying and/or increasing modes within the $\Delta T$
instability domain. But as soon as $T > T_{+}$ the waves become
stable again and their amplitudes do not change anymore The
resulting wave amplitude is enhanced/diminished in comparison with
the initial amplitude depending whether transiently
increasing/decreasing component was dominant for the initially
excited wave.

Numerical results, represented by Figs. 2-5, fully confirm these
qualitative expectations. Figs. 2 and 3 illustrate the inversion
symmetry of the functions ${\Psi}_a(T)$. The first of these two
plots shows the temporal evolution for $a=0.1$ and with initial
conditions ${\Psi}_a(-100)=0.01$ and ${\Psi}_a^{(1)}(-100)=0$. The
figure is plotted for the interval $-100<T<100$. Evidently the
inversion symmetry implies that an another solution of this
equation for the same value of $a$ but with the initial values
${\hat{\Psi}}_a(-100)={\Psi}_a(100)$ and
${\hat{\Psi}}_a^{(1)}(-100)=-{\Psi}_a^{(1)}(100)$ will be exactly
inverse-symmetric. In this particular example these initial values
are ${\hat{\Psi}}_a(-100)=-0.1354$ and
${\hat{\Psi}}_a^{(1)}(-100)=-0.2186$. The inversion symmetry of
these solutions is apparent. Physically this fact implies that the
presence of the shear flow ensures burst-like and robust increase
of amplitudes (energy) of some Alfv\'en waves, while there are
always other waves which, on the contrary, sharply loose their
energy under the influence of the shear flow.

Momentary appearance of the transient instability on the presented
graphs is related to the narrowness of the transient instability
interval, which is apparent from Fig.1. Note that the plotting
time interval on Fig.1 is taken very narrow in order to give
magnified portrait of the behavior of ${\Omega}_{\rm eff}$, while
Figs.2-3 are deliberately drawn for the much wider range in order
to illustrate the behavior of waves on a large time span.

The family of solutions ${\Psi}_a(T)$ of (31) does not represent a
physical variable of the problem, so in order to recover
information about the temporal evolution of perturbations for
physical variables it is more convenient to solve numerically
(22-23) and to recover components of vectors ${\bf v}$ and ${\bf
b}$ from (24). Besides, it is instructive to calculate the kinetic
energy, the magnetic energy and the total energy of perturbations
as given by (25). In order to track the temporal evolution of the
shear-modified wave branch as such we set the amplitude of the
unmodified component ${\cal B}(t)$ to be zero ${\cal C}=0$.
Alternatively our numerical task was, first, to solve Eq.(28) in
order to get functions ${\Psi}(t)$ and ${\Psi}^{(1)}(t)$. Second
step was to calculate the physical variables by means of
Eqs.(22-25).

This set of calculations was performed for different values of the
system parameters and some representative examples are given in
Figs. 4-5. Note that they are plotted as functions of the real
time variable $t$ and not the variable $T$ used in (31). From
these figures we readily see that when $\Delta = 0$
($k_{\perp}=0$), the shear flow efficiently ``pumps" energy into
the longitudinal components of the velocity and the magnetic field
perturbations, while the transverse components stay basically
unchanged aside from the transitory, ``burst-like" increase of
their amplitudes in the brief, transient amplification phase. This
is another indication of the above-mentioned fact that the
velocity shear primarily affects the incompressible limit of the
fast magnetosonic waves. The asymptotic increase of the total
energy, as shown in Fig.4, being entirely due to the increase in
$v_z$ and $b_z$, is quite substantial (about two orders of
magnitude for the given example).

\begin{figure}
  \resizebox{\hsize}{!}{\includegraphics[angle=90]{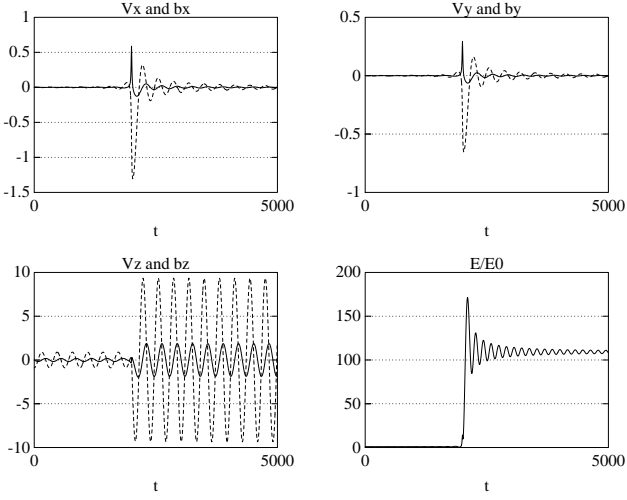}}
  \caption{The numerical solution of Eq.(28) presented for all
components of ${\bf v}$ (solid lines on first three plots) and
${\bf b}$ (dashed lines on first three plots)  vectors. Last plot
shows time evolution of $E_{\rm tot}(t)/E_{\rm tot}(0)$. The
values of parameters are: $C_{\rm A}=0.2$, $k_x=20$, $k_y=10$,
$k_z=0.1$, $C_x=0.1$, $C_y=0.05$, $a=0.1789$. Initial values are:
${\Psi}(0) =0.01$ and ${\Psi}^{(1)}(0)=0$.}\label{fig4}
\end{figure}

\begin{figure}
  \resizebox{\hsize}{!}{\includegraphics[angle=90]{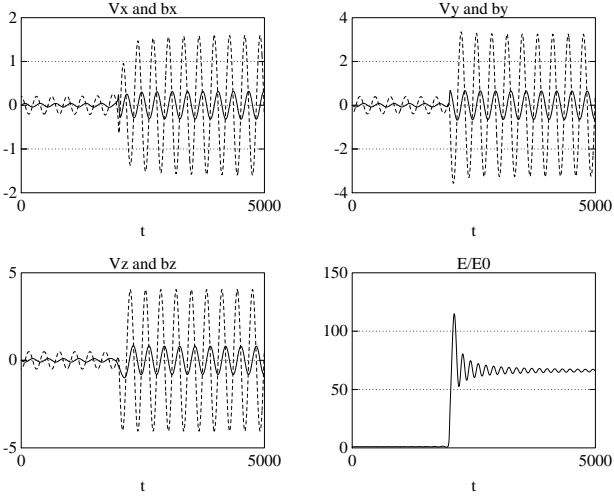}}
  \caption{The numerical solution of Eq.(22-23) with the same
parameters as on Fig.4 except $k_y=10.1$.}\label{fig5}
\end{figure}

This is a typical mode of behavior for perturbations with
${\Delta}=0$ and it is quite similar with the behavior of
hydromagnetic waves in plane shear flows (\cite{cckl93}). By
imparting a small but nonzero ${\Delta}$ (see Fig. 5), we can make
all perturbation components grow. However, the overall increase of
the total energy of perturbations in the latter case is somewhat
smaller than in the former case.

\section{Helical flow}

For more complicated MHD flows, the time evolution of the
wavenumber vector ${\bf k}(t)$, governed by equations (14),
becomes {\it nonlinear} and it makes the temporal behavior of
linear perturbations much more complex. A relatively simple, 2-D
hydrodynamic system of such complexity was recently investigated
(\cite{mr99}). For the MHD {\it helical flow}, specified by the
shear matrix (11) and considered in this section, we may deduce
from (15) and (13c) that the  perturbations may  grow either
exponentially (when differential rotation rate $n>1$), or vary in
time in a periodic way (when $n<1$). Below we develop the
mathematical formalism for the study of perturbation evolution for
these two classes of solutions.

\subsection{General formalism}

The starting set of equations is again (16-19).  Taking time
derivative of (16) we find out that
$$
({\bf k}{\cdot}{\bf
v})^{(1)}=-({\cal S}^T{\cdot}{\bf k}){\cdot}{\bf v}-{\bf k}{\cdot}
({\cal S}{\cdot}{\bf v})-|{\bf k}|^2(P+C_{\rm A}^2b_z)=0,
$$
which after
noticing that $({\cal S}^T{\cdot}{\bf k}){\cdot}{\bf v}= {\bf
k}{\cdot}({\cal S}{\cdot}{\bf v})$ implies that the following
algebraic relation holds:
$$
P+C_{\rm A}^2b_z=-2{\bf k}{\cdot}({\cal
S}{\cdot}{\bf V})/|{\bf k}|^2=2{\bf k}^{(1)} {\cdot}{\bf v}/|{\bf
k}|^2. \eqno(33)
$$

Note that this relation follows from (16), which, in turn, is direct
consequence of the `incompressibility' condition. The
relation (33), applied to (17), helps to reduce the initial
system to the following pair of the first-order ODE's:
$$
{\bf v}^{(1)}+{\cal S}{\cdot}{\bf v}=C_{\rm A}^2k_z{\bf b}-2({\bf
k}^{(1)}{\cdot}{\bf v}) {\bf k}/|{\bf k}|^2, \eqno(34)
$$
$$
{\bf b}^{(1)}={\cal S}{\cdot}{\bf b}-k_z{\bf v}. \eqno(35)
$$

It is worthwhile also to note that $ {\bf k}({\bf
k}^{(1)}{\cdot}{\bf v})={\bf v}{\times}({\bf k}{\times}{\bf
k}^{(1)})$, which allows to rewrite (34) in the following form:
$$
{\bf v}^{(1)}+{\cal S}{\cdot}{\bf v}=C_{\rm A}^2k_z{\bf b} +{\bf
R}{\times}{\bf v}, \eqno(36)
$$
with ${\bf R}$ defined as:
$$
{\bf R}(t){\equiv}({\bf k}{\times}{\bf k}^{(1)})/|{\bf k}|^2.
\eqno(37)
$$
The interesting property of this form of the Eq.(34) is that the
only time-dependent coefficient is just represented by one vector
$R(t)$!

We can reduce (34--35) to the following explicit second-order
equation for the magnetic field:
$$
{\bf b}^{(2)}+{\left[{\omega}_{\rm A}^2-{\cal S}^2 \right]}{\bf b}+{{2{\bf
k}}\over{|{\bf k}|^2}} {\left[{\bf k}^{(1)}{\cdot}{\bf
b}^{(1)}+{\bf k}^{(2)}{\cdot}{\bf b} \right]}=0. \eqno(38)
$$

This equation is quite informative, because it shows explicitly the effects
of the helical flow imposed upon the evolution of perturbations. Note that the term
containing ${\bf k}^{(1)}{\cdot}{\bf b}^{(1)}$
is also present in the pure outflow case (see for comparison Eq. (27)). While terms containing
${\cal S}^2{\bf b}$ and ${\bf k}^{(2)} \cdot {\bf b}$ are characteristic to the case of helical
flow and they reflect different aspects of shear-induced
processes in helical flows.

Obviously (38) is a general equation and we can base our analysis
on it. However a different approach may also be useful. Let us
introduce ``projection variables":
$$ U{\equiv}{\cal
S}_{ij}k_iv_j=-{\bf k}^{(1)}{\cdot}{\bf v}, \eqno(39a)
$$
$$
H{\equiv}{\cal S}_{ij}k_ib_j=-{\bf k}^{(1)}{\cdot}{\bf b},
\eqno(39b)
$$
$$ V{\equiv}{\cal S}^2_{ij}k_iv_j={\bf
k}^{(2)}{\cdot}{\bf v}, \eqno(39c)
$$
$$
B{\equiv}{\cal
S}^2_{ij}k_ib_j={\bf k}^{(2)}{\cdot}{\bf b}. \eqno(39d)
$$

Using these definitions and our basic set of equations we find:
$$
U^{(1)}=-2V+C_{\rm A}^2k_zH-{{2({\bf k}{\cdot}{\bf
k}^{(1)})}\over{|{\bf k}|^2}}U, \eqno(40)
$$
$$
H^{(1)}=-k_zU, \eqno(41)
$$
$$
V^{(1)}=-2{\Gamma}^2U+C_{\rm A}^2k_zB+{{2({\bf
k}{\cdot}{\bf k}^{(2)})}\over{|{\bf k}|^2}}U, \eqno(42)
$$
$$
B^{(1)}=-k_zV. \eqno(43)
$$

From this set we can derive the following two second-order equations
for $H$ and $B$:
$$
H^{(2)}+{{2({\bf k}{\cdot}{\bf
k}^{(1)})}\over{|{\bf k}|^2}}H^{(1)}+{\omega}_{\rm A}^2H= -2B^{(1)},
\eqno(44)
$$
$$
B^{(2)}+{\omega}_{\rm A}^2B=f(t)H^{(1)}, \eqno(45)
$$
where
$$
f(t){\equiv}{{2({\bf k}{\cdot}{\bf k}^{(2)})}\over{|{\bf
k}|^2}}-2{\Gamma}^2. \eqno(46)
$$

Moreover, if we define
$$
{\Psi}{\equiv}|{\bf k}|H, \eqno(47)
$$
we can rewrite (44) also as:
$$
{\Psi}^{(2)}+{\left[{\omega}_{\rm A}^2-|\bf R|^2-{{({\bf k}{\cdot}{\bf
k}^{(2)})}\over {|{\bf k}|^2}}\right]}{\Psi}=-2|{\bf k}|B^{(1)}.
\eqno(48)
$$

Note the appearance of the vector ${\bf R}$ in this equation.
Overall, (45) and (48) give rather
transparent analytic presentation, which can be used for the
analysis of interesting classes of solutions. Note also that in
the absence of rotation $f(t)=0$ these equations become decoupled
and they reduce to the ``pure outflow" case. In that case, we had
a splitting of the mode into the usual Alfv\'en and shear-modified
(``hybrid" as we can also call it) Alfv\'en mode. Below we shall see
that rotation not only gives birth to new instabilities but it
couples these two branches of the Alfv\'en mode!

\subsection{Exponential ${\bf k}(t)$ case}

Depending on the actual type of the motion in the plane normal to
$Z$ there will be different regimes of the time variation of ${\bf
k}(t)$ (see for details \cite{mr99}). Instabilities are present
both when the wavenumber varies exponentially and periodically.

In the former case, i.e., when the absolute value of ${\bf k}(t)$
increases exponentially, the dissipation effects will
eventually start to be important and try to damp the
mode (\cite{mr99}) despite the presence of the shear instability.
The same argument is valid for those cases,
too, when the temporal growth of the wavenumber is linear. Therefore,
considering shear-induced nonmodal effects in inviscid fluids in cases
when $|{\bf k}|$ is monotonously increasing, we should realize that
the description is physically meaningful only for initial (finite)
times.

\begin{figure}
  \resizebox{\hsize}{!}{\includegraphics[angle=90]{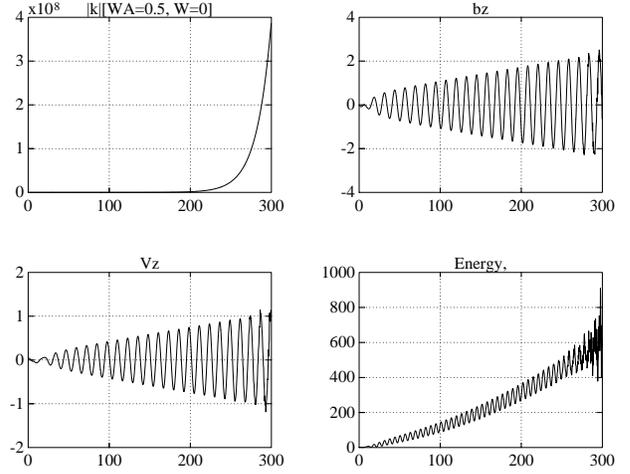}}
  \caption{The numerical solution for the case when temporal evolution of
  ${\bf k}$'s is exponential. The values of parameters are: $k_x=5$, $k_y=10$,
  $k_z=1$, $\sigma = 10^{-2}$, $A_1=0.3$, $A_2=10^{-2}$, $C_1=0.8$, $C_2=0.9$,
  $C_{\rm A}=0.5$.}\label{fig6}
\end{figure}

It is worthwhile to make a look at some of these solutions. On
Fig.6 we present the results of numerical calculations, taken for
the case, when $k_x=5$, $k_y=10$, $k_z=1$, $\sigma = 10^{-2}$,
$A_1=0.3$, $A_2=10^{-2}$, $C_1=0.8$, $C_2=0.9$, $C_{\rm A}=0.5$. The
value of ${\Gamma}^2=3.1{\times}10^{-3}$ is positive and the value
of $|{\bf k}(t)|$ is exponentially increasing (see Fig.6a). The
characteristic amplification time scale ${\cal
T}{\simeq}{\Omega}^{-1}$ is of the order of  the period of
rotation. The resulting wave evolves accordingly, which is clearly
visible from the figures. It is noteworthy to see that, analogous
to the pure outflow case, shear flow energy is most efficiently
absorbed by the longitudinal components of the velocity and the
magnetic field perturbations. Still, now, the energy growth,
plotted again for the $E_{\rm tot}(t)/E_{\rm tot}(0)$, is not algebraic
but exponential and is entirely due to the exponential evolution
of the perturbation wave number vector.

\subsection{Periodic ${\bf k}(t)$ case}

When values of ${\Gamma}^2$ are negative, the temporal evolution
of ${\bf k}(t)$'s is periodic, which means that the absolute value
of the wavenumber vector stays bounded within certain limits. It
means that in this case shear-induced effects have potentially
much wider durability and may potentially lead to strongly
perceptible effects. The limits of this paper does not allow us to
present full analysis of all possible classes of solutions. The
most interesting feature of this case is that Alfv\'en waves
become parametrically unstable in a helical flow. The Figs.7-9
demonstrate this noteworthy fact. All figures are drawn for the
case when $A_1=-A_2=0.3$, and ${\sigma}=0.01$, so that
$W{\equiv}(-{\Gamma}^2)^{1/2}{\simeq}0.2998$. Fig.7 shows the case
when  Alfv\'en frequency is ${\omega}_{\rm A}=0.29$. The system is
still out of the parametric resonance and the system, therefore,
exhibits just Alfv\'en waves, modulated by the periodic shear.
The picture is drastically different when  ${\omega}_{\rm A}=0.3$
--- the Fig.8 shows that there appears a robust instability of
parametric (resonant) nature that leads to the strong exponential
increase of the system energy in the same span of time as on
Fig.7. The resonance is quite sharp, because as soon as Alfv\'en
frequency slightly exceeds the value of $W$ the system becomes
stable again, as it is visible from the Fig.9, which is drawn for
the case when ${\omega}_{\rm A}=0.31$.

\begin{figure}
  \resizebox{\hsize}{!}{\includegraphics[angle=90]{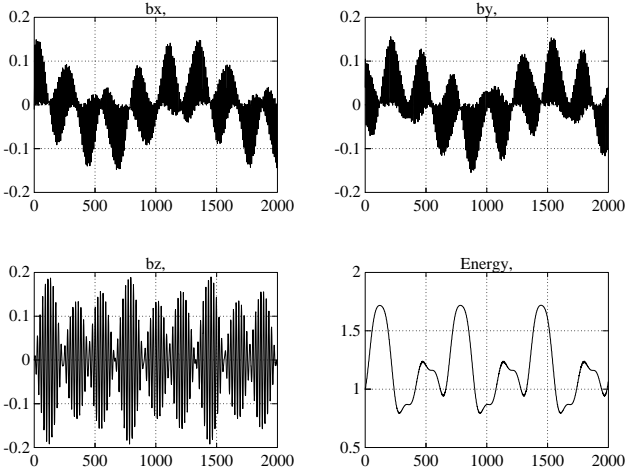}}
  \caption{The numerical solution for the case when
  $A_1=-A_2=0.3$, ${\sigma}=0.01$, $C_1=0.8$, $C_2=0.9$,
   $W{\simeq}0.2998$ and ${\omega}_{\rm A}=0.29$.}\label{fig7}
\end{figure}

\begin{figure}
  \resizebox{\hsize}{!}{\includegraphics[angle=90]{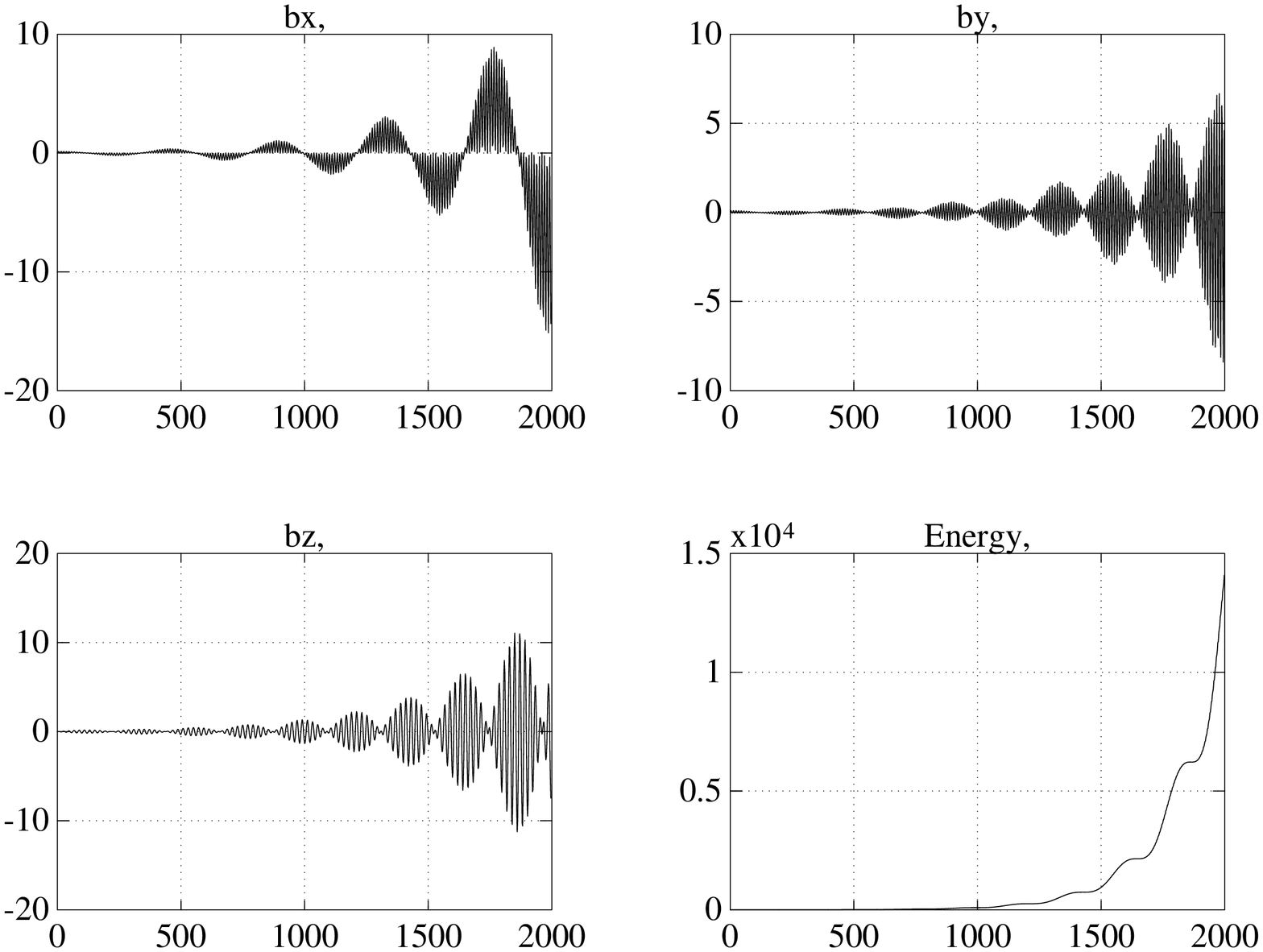}}
  \caption{The numerical solution for the case when
$W{\simeq}0.2998$ (the same set of parameters as on the Fig.7) but
with ${\omega}_{\rm A}=0.3$.}\label{fig8}
\end{figure}

\begin{figure}
  \resizebox{\hsize}{!}{\includegraphics[angle=90]{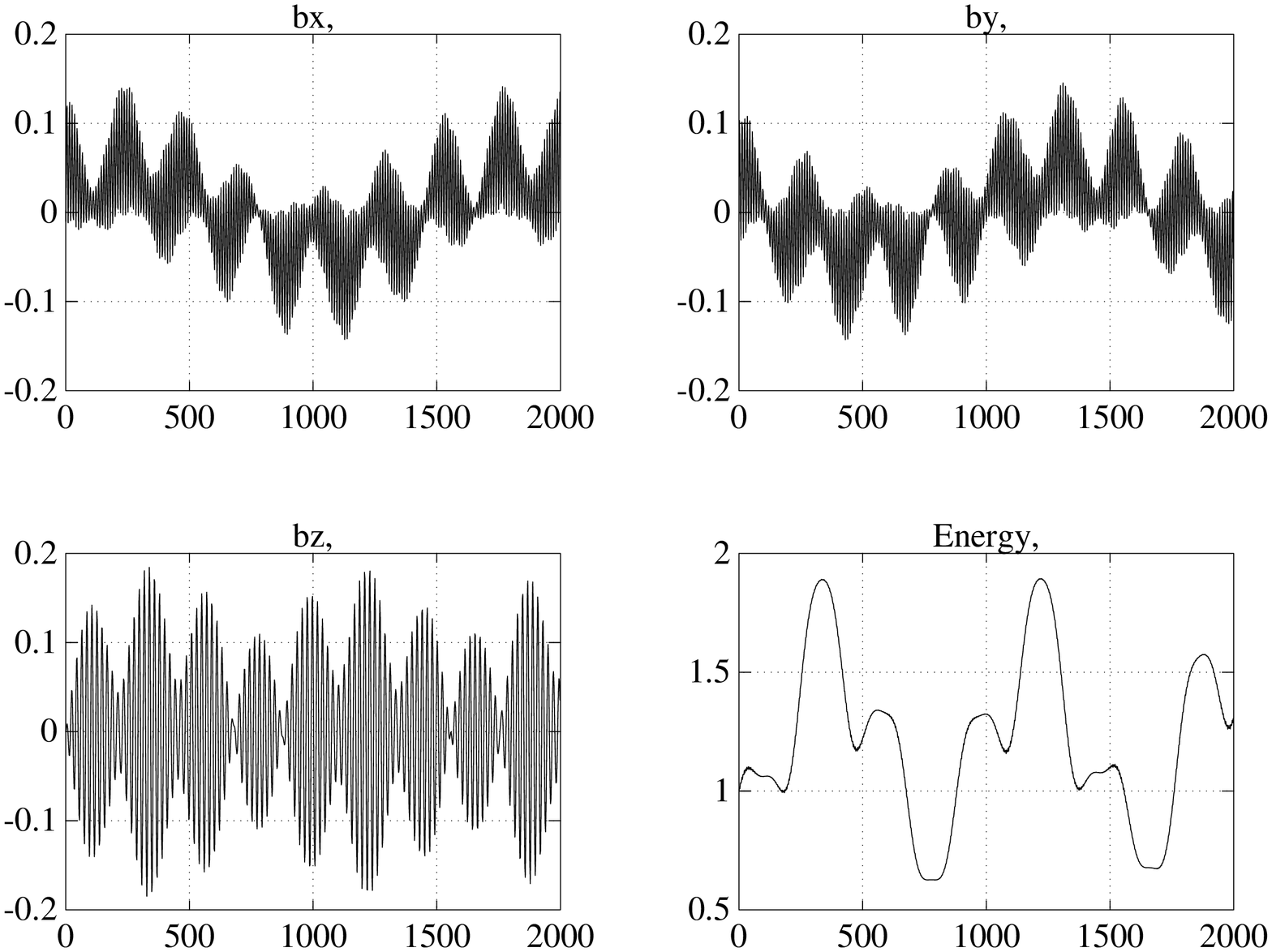}}
  \caption{The numerical solution for the case when
$W{\simeq}0.2998$  (the same set of parameters as on the Fig.7)
but with ${\omega}_{\rm A}=0.31$.}\label{fig9}
\end{figure}

\begin{figure}
  \resizebox{\hsize}{!}{\includegraphics[angle=90]{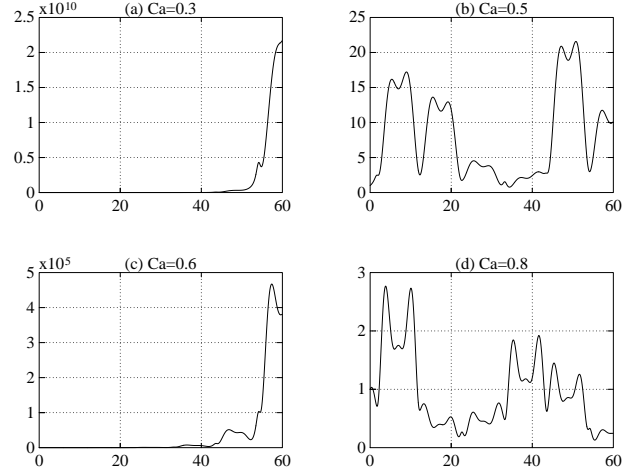}}
  \caption{The numerical solution for total energy of perturbations
  normalized on their initial values. The values of parameters are: $k_x=20$,
  $k_y=5$, $k_z=1$, $\sigma = 0.6$,
$A_1=0.5$, $A_2=-0.9$, $C_x=0.01$, $C_y=0.05$. In this case
$W=0.3$, while ${\omega}_{\rm A}=0.3, 0.5, 0.6,$ and $0.8$ on the
plots labeled (a), (b), (c), and (d), respectively.}\label{fig10}
\end{figure}

\begin{figure}
  \resizebox{\hsize}{!}{\includegraphics[angle=90]{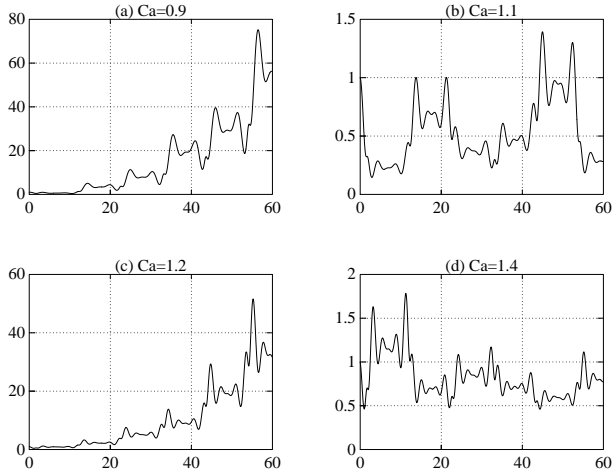}}
  \caption{The numerical solution for the total energy of
  perturbations. The values of parameters are the same as on
  Fig.10. while ${\omega}_{\rm A}=0.9, 1.1, 1.2,$ and $1.4$ on the plots
labeled (a), (b), (c), and (d), respectively.}\label{fig11}
\end{figure}

Further study shows that actually there are several isolated
regions of instability, which happen to be centered around the
values ${\omega}_{\rm A}=nW, n=1,2,..$ and usually first resonance
$W={\omega}_{\rm A}$ is the most efficient one. This is a strong
indication in favor of the resonant and parametric nature of this
instability. Fig.10 and Fig.11 illustrate this
interesting property of the system. The plots are drawn for the
total energy of a perturbation, normalized on its initial value:
$E_{\rm tot}(t)/E_{\rm tot}(0)$. The simulation is made for
the case when $k_x=20$, $k_y=5$, $k_z=1$, $\sigma = 0.6$,
$A_1=0.5$, $A_2=-0.9$, $C_x=0.01$, $C_y=0.05$. Note that in this
case $W=0.3$. Making simulations for different values of the
Alfv\'en velocity we see that perturbations are strongly unstable
when ${\omega}_{\rm A}{\simeq}nW$. Namely the instability is present on
Fig.10 (a) and (c) and Fig.11 (a) and (c), where the values of
${\omega}_{\rm A}$ are $0.3$, $0.6$, $0.9$, and $1.2$, respectively. At
the same time we see that for intermediate values of ${\omega}_{\rm A}$,
when $nW<{\omega}_{\rm A}<(n+1)W$, the Alfv\'en waves stay stable.
Unfortunately the complexity of the system does not allow us to
perform strict analytic analysis and locate the actual width and
size of instability regions. However, our results {\it do} show that
there are several such regions and presumably the efficiency of
the instability decreases with the increase of $n$.

\section{Discussion}

The results of the present study clearly show that both helical
and purely ejectional flows of MHD plasmas could serve as
efficient ``amplifiers" of Alfv\'en waves. In alliance with other
shear-induced phenomena (such as shear-induced wave
transformations and/or generation of shear vortices), these
processes could lead to the cascade amplification of hydromagnetic
waves leading to MHD turbulence in these  flows (\cite{rpm00}). In
more general terms  it could be stated that the presence of the
velocity shear strongly modifies the nature of Alfv\'en  waves
such that they can extract energy from the mean flow.

Certainly, while talking about these processes, we should bear in
mind that due to the limitation of the `nonmodal approach' we are
not able to describe quantitatively the onset of these processes
in the real physical space. Besides the linearized nature of our
calculations prevents us from  taking into account of the back
reaction of shear-modified Alfv\'en waves on the mean flow. The
last but not the least limitation is related with the neglect of
the dissipation in the flow.

Can Alfv\'enic perturbations still extract energy from the
background flow when dissipation is present? For magnetoacoustic
modes sustained by a simple, linearly sheared two-dimensional flow
it was recently found (\cite{bprr01}) that the presence of
dissipation doesn't prevent a shear-modified MHD mode from the
energy extraction. For the case of helical dissipative flows until
now no numerical studies were done. Numerical study of these
processes in the swirling flows requires 3-dimensional MHD
simulations over large time intervals, which is quite a tough
challenge. We expect that the outcome of the ``competition"
between the increase of the energy due to the presence of the
shear flow and viscous damping due to the presence of  the
dissipation will depend on the time scales of both processes and
initial conditions. Relevant numerical studies are planned to be
started in the near future.

However, basing on our results about `nonmodal' temporal evolution
of Alfv\'enic perturbations, we can give qualitative picture of
the physical appearance of these Alfv\'en waves in the presence of
viscous dissipation for all those three subclasses of flows --
viz: (a) ejectional outflows with no rotation; (b) helical flows
with exponentially increasing $|{\bf k}(t)|$; (c) helical flows
with periodically changing $|{\bf k}(t)|$) -- which were
considered.

For an ejectional outflow, with linear temporal growth of the
$|{\bf k}(t)|$, a monochromatic Alfv\'enic disturbance would
undergo {\it one} act of transient, `burst-like' amplification
followed by the viscous decay. For the package, containing
harmonics of different $|{\bf k}(0)|$, one could expect a sequence
(`cascade') of their transient amplifications mixed with their
viscous damping. This sort of the Alfv\'en wave, while passing
through the flow,  would slowly decay but repeatedly regain its
energy via the chain-sequence of amplification events.

For helical flows, where the temporal increase of the  $|{\bf
k}(t)|$ is exponential, the Alfv\'en waves could be less fortunate.
It is true that their energy might grow exponentially, but the
increase of the  $|{\bf k}(t)|$ seems to be a more rapid process
than the increase of the total energy of the Alfv\'en wave (see
Fig.6). Therefore, even if initially the Alfv\'en waves could
extract some energy from the background flow, the powerful viscous
damping would sooner or later prevail and return the energy back
to the flow in the form of heat. It means that in this situation,
perhaps, one could not see any high amplitude Alfv\'en waves, but
one would rather witness an efficient heating of the flow due to
this {\it `self-heating'} process: when Alfv\'en waves serve as energy
transmitters, extract a part of the flow's regular kinetic energy
and returning the energy back to the flow in the form of the heat.

For helical flows, where the variability of the  $|{\bf k}(t)|$ is
periodic, the situation can be the most exotic. The strength of the
viscous damping in such flows is limited, because the value of the
$|{\bf k}(t)|$ is kept finite: it doesn't increase neither
algebraically nor exponentially. At the same time, Alfv\'en waves
are modulated by the presence of the flow and there exist separate
regions of parametric instability, which implies that under
favourable conditions the Alfv\'en waves propagating through this
kind of flow can be unstable. Therefore, flow patterns of this
gender might serve as efficient amplifiers of Alfv\'en waves.

In all these cases the ultimate `fate' of any individual Alfv\'en
wave would depend on the time- and length-scales of their
propagation and the flow geometry, as well as, of course, on the
actual strength of the viscous dissipation in the flow.

The shear-modified Alfv\'en waves seem to be the {\it hybrids} of
the usual Alfv\'en waves and the so called `Kelvin modes'
(\cite{mp77, rcm98}). The latter are nonperiodic, transiently
growing modes inherent to the incompressible neutral shear flows.
In the absence of the equilibrium magnetic field, the Kelvin
Transients are the only modes of collective behaviour exhibited by
the flow.  The shear-modified Alfv\'en waves are formed by
combining the features of the usual waves with the transient shear
vortices; their ability to extract/give energy from/to the mean
flow follows from the latter.

Within the framework of the nonmodal approach we are able to track
and study properties of these shear-modified Alfv\'en waves in the
space of wave numbers ($\bf k$-space). The orientation and
magnitude of a wave number vector, being variable in time
and governed by Eq. (14), varies because of the local straining of
the background flow. The evolution of the ${\bf k}(t)$ field, in
its turn, exercises influence  on the temporal behaviour of
physical variables: density and pressure perturbations, components
of the magnetic field and the velocity vectors. In terms of the
usual physical coordinates (space and time) these variables are
periodic in each space coordinate. However,  these modes are
`non-normal' (\cite{bf92}) because both wavenumbers and amplitudes
associated with each disturbance are functions of time, implying
that the solutions are {\it not} separable in space and time. The
latter, nonexponential time-dependence of the solutions is
disclosed through the nonmodal approach and it helps to
``smoke-out" new classes of modes hardly accessible through
traditional normal mode analysis. In order to have complete
real-space physical picture of the evolution of these Alfv\'enic
perturbations we need, therefore, to develop phenomenological
approach and try to understand the dynamics in terms of vorticity
dynamics (\cite{cc86}). Alternatively one might approach the
problem through direct numerical simulations, generalizing the
very first attempts of numerical studies of much simpler MHD
velocity patterns (\cite{bprr01}).

Turning our attention back to astrophysics, we might argue that
the role of the shear-modified hybrid Alfv\'en waves
could be quite important in a number of astrophysical situations.
Let us first assess the relevance these waves could have to the
{\it solar plasma flows}. It is known that higher layers of the
solar chromosphere consist of long, vertical columns ({\it
spicules}) rising above the general background (\cite{a86}) and
well-visible at the extreme limb of the Sun, where the overall
structure of the chromosphere resembles a ``burning prairie."  The
spicules rise vertically, like slender magnetic flux tubes
(filaments) out of the network beginning in the chromosphere and
threading through the solar transition region into the low corona.
Their average lifetime is 10-15 minutes, characteristic length
scale $10.000 km$, and characteristic diameter $700-1000 km$.
Often the spicules appear to have an ``Eifel tower" shape with
vertical flow velocities in the range
 $20-25 km/s$. There is also some evidence for plasma
rotation in spicules (\cite{ks99}). It means that the plasma
motion within spicules may easily be of helical nature. The recent
discovery of swirling macrospicules, called {\it solar tornados}
(\cite{pm98}) provides additional evidence in this direction.

If we imagine flow patterns, like spicules or solar tornadoes,
with moderate storage of total (quasi)equilibrium energy it may
happen that transiently amplified Alfv\'en waves, bred within
these structures, could efficiently `suck out' much of their
energy and lead to eventual exhaustion and disappearance of the
flow patterns. The shear-induced cascade of Alfv\'en waves could
take away much of the available energy of the ``parent" flow. The
observational consequences may be dramatic; we could imagine
individual spicules acting as short-lived plasma cannons firing
away heavy blasts of large-amplitude Alfv\'en waves and
disappearing after the wave package has drained most of the energy
from the ``parent" flow.

It is quite fascinating to surmise that cascade of transient
amplifications of Alfv\'en waves within spicules, together with
other kinds of shear-induced wave processes (like, e.g., ``MHD
wave oscillations" \cite{rpm00}) may account for the quasisteady
appearance of spicules in the solar atmosphere. Taking the
``burning prairie" metaphor a little further,  we may speculate
that shear induced processes, ``burning"  individual ``grass blades",
could be responsible for the overall conflagration of the spicule
network which, in turn, provides an efficient mechanism for the
transfer of energy from the quasisteady motions of solar plasmas
into the flux of large-amplitude Alfv\'en waves. These waves are
able to penetrate through the solar transition region and reach
the lower layers of the solar corona. Here the processes of the
shear-induced ``MHD wave oscillations" (\cite{rpm00}) may
contribute to the appearance of compressible MHD modes (slow
and/or fast magnetosonic waves) which, in turn, are effectively
damped giving back their energy to the solar plasma particles,
accelerating them and providing initial launch of the solar wind
flows. This highly speculative picture must be  carefully
developed and tested in order to make it convincing. We are in the
process of developing a numerical model for an individual helical
solar flow pattern, and hope to see the above-described processes
in  real physical space through  numerical simulations.

The parametric Alfv\'en instability  may be effective in flows,
where conditions are favorable for the
correlation between the rate of rotation and stretching of the
flow lines (which determines the value of the characteristic
frequency $W$) and the strength of the equilibrium magnetic field
(which determines the magnitude of the Alfv\'en frequency
${\omega}_{\rm A}$). Such a correlation may easily happen within solar
jet flows, where both the ejectional and rotational modes of
motion are present. Any time such correlation takes place it will
lead to a burst-like generation of high-amplitude Alfv\'en waves
with corresponding ``exhaustion" (and, maybe, even disappearance)
of the ``parent" helical flow pattern. Since the modern apparatus
used on the last generation solar satellites enables one to
measure both the strength of magnetic fields and the rate of
rotation, one can assume that it will be fairly realistic to
verify whether the appearance of high-amplitude Alfv\'en waves is
really related to shear-induced resonant parametric instability of
helical flows.

We think that shear-induced instabilities of this nature may be
also present in
{\it accretion-ejection flows}.  The differential rotation
parameter $n$ controls the ${\bf k}(t)$  dynamics which, in turn,
specifies the nature of fluctuations in a given flow. When $n<1$
(including the rigid rotation case) $\Gamma$ is imaginary and
${\bf k}(t)$ is periodic. In such cases (e.g., innermost regions
of galactic gaseous disk, believed to have almost constant
rotation rates) the system may sustain parametrically unstable
Alfv\'en waves. While when $n>1$ (including the Keplerian rotation
regime), $\Gamma$ is real and makes the time behavior of ${\bf
k}(t)$ exponential. This regime can be realized in different kinds
of quasi Keplerian accretion disks. Note that the characteristic
amplification time scale ${\cal T}{\simeq}{\Omega}^{-1}$ is of the
order of  the inverse period of rotation, being at least of the
same order as the ``Magnetorotational instability" growth rate
(\cite{bh98}). Since the latter instability is thought to be the
strongest one, accounting for the turbulence in accretion disks,
we could surmise that the shear-induced instability discussed  in
this paper could also play a role in the onset of  turbulence
in accretion-ejection flows.

Finally, considering our results in the context of galactic gaseous disks
and their large-scale magnetic fields, we might envisage a
possible relation between the shear-induced amplification of
Alfv\'en waves and standard dynamo action in swirling flows.
Within galactic disks algebraic and/or usual instabilities of the
nonmodal origin might provide the important mechanism for the
initial amplification of the magnetic energy and this process
could compete with the omnipresent Ohmic diffusion. This is a
challenging task to see what is  the role of nonmodal,
shear-induced processes in the framework of the general theory of
dynamo generation of galactic magnetic fields.

\section{Acknowledgements}

Gianluigi Bodo, Silvano Massaglia and Andria Rogava  were
supported, in part, by the INTAS grant No.~97-0504. Andria Rogava
is grateful to the Abdus Salam I.C.T.P. and Universit\'a degli
Studi di Torino for supporting him, in part, through the Regular
Associate Membership Award and the {\it Assegno di Ricerca},
respectively.

{}


\begin{thebibliography}{}


\bibitem[Athay 1986]{a86} Athay R.G. 1986, in {\it Physics of the Sun}, Vol. II, ed. P. A.
Sturrock (Dordrecht: Reidel), 51.

\bibitem[Balbus \& Hawley 1991]{bh91} Balbus S. A.,  \& Hawley J. F.
1991, ApJ, 376, 214

\bibitem[Balbus \& Hawley 1998]{bh98} Balbus S.A., \& Hawley J.F. 1998, Rev. of Mod. Phys., 70, 1

\bibitem[Balogh et al. 1995]{bst95} Balogh A., Smith E.J., Tsuritani B.T., et al. 1995,
Science, 268, 1007

\bibitem[Belien et al. 2002]{bbgvk02} Belien A. J. C., Botchev M.
A., Goedbloed J. P., van der Holst B., \& Keppens R. 2002, Comp.
Phys. Comm., 147, 497


\bibitem[Bodo et al. 2001]{bprr01} Bodo G., Poedts S., Rogava A.D., \& Rossi P.
2001, A\&A, 374, 337

\bibitem[Butler \& Farrell 1992]{bf92} Butler K.M., \& Farrell B.F. 1992, Phys. Fluids A, 4, 1637

\bibitem[Chagelishvili et al. 1993]{cckl93} Chagelishvili G.D., Chanishvili R.G.,
Khristov T.S., \& Lominadze J.G. 1993, Phys. Rev. E, 47, 366


\bibitem[Chagelishvili et al. 1996]{crt96} Chagelishvili G.D., Rogava A.D., \& Tsiklauri
D.G. 1996,  Phys. Rev. E, 53, 6028

\bibitem[Chagelishvili et al. 1997]{crt97} Chagelishvili G.D., Rogava A.D., \& Tsiklauri
D.G. 1997, Phys. Plasmas, 4, 1182



\bibitem[Craik \& Criminale 1986]{cc86} Craik A.D.D., \& Criminale W.O. 1986, Proc. R. Soc.
Lond. A, 406, 13

\bibitem[Ferrari 1998]{f98} Ferrari A. 1998, ARA\&A, 38, 539

\bibitem[Hollweg 1990]{h90} Hollweg J. V. 1990, Comp. Phys. Reports, 12, 205

\bibitem[Kudoh \& Shibata 1999]{ks99} Kudoh T., \& Shibata K. 1999, ApJ, 514, 493

\bibitem[Lagnado et al. 1984]{l84} Lagnado R.R., Phan-Thien N., \& Leal L.G. 1984, Phys.
Fluids, 27, 1094


\bibitem[Mahajan et al. 1997]{mmr97} Mahajan S.M., Machabeli G.Z., \& Rogava A.D. 1997,
ApJ, 479, L129

\bibitem[Mahajan \& Rogava 1999]{mr99} Mahajan S.M., \& Rogava A.D. 1999, ApJ, 518, 814

\bibitem[Marcus \& Press 1977]{mp77} Marcus P., \& Press W. H. 1977, J. Fluid Mech., 79, 525

\bibitem[Mikhailenko et al. 2000]{mms00} Mikhailenko V. S., Mikhailenko V. V., \& Stepanov K. N.
2000, Phys. Plasmas, 7, 94

\bibitem[Peter 2001]{p01} Peter H. 2001, A\&A, 374, 1108

\bibitem[Pike \& Mason 1998]{pm98} Pike C.D., \& Mason H.E. 1998, Solar Phys., 182, 333

\bibitem[Poedts et al. 1998]{prm98} Poedts S., Rogava A.D., \& Mahajan S.M. 1998, ApJ, 505, 369


\bibitem[Poedts et al. 2000]{pkr00} Poedts S., Khujadze G.R., \& Rogava A.D. 2000,  Phys.
Plasmas, 7, 3204

\bibitem[Prikryl et al. 2002]{ppmy02} Prikryl P., Provan G.,
McWilliams K.A., \& Yeoman T.K. 2002, Annales Geophysicae, 20, 161

\bibitem[Rogava et al. 1996]{rmb96} Rogava A.D., Mahajan S.M., \& Berezhiani V.I.
1996, Phys. Plasmas, 3, 3545

\bibitem[Rogava \& Mahajan 1997]{rm97} Rogava A.D., \& Mahajan S.M. 1997, Phys. Rev. E, 55, 1185


\bibitem[Rogava et al. 1998]{rcm98} Rogava A.D;, Chagelishvili G.D., \& Mahajan S.M. 1998,
Phys. Rev. E, 57, 7103

\bibitem[Rogava \& Poedts 1999]{rp99} Rogava A.D., \& Poedts S., in Proc. of the 9th European
Meeting on Solar Physics (Firenze, Italy, September 12-18, 1999),
Ed. A. Wilson, ESA Sp-448, (1999).

\bibitem[Rogava et al. 1999]{rph99} Rogava A.D., Poedts S., \& Heirman S. 1999, Mon. Not. R.
Astron. Soc., 307, L31

\bibitem[Rogava et al. 2000]{rpm00} Rogava A.D., Poedts S., \& Mahajan S.M. 2000, A\&A, 354, 749

\bibitem[Ryutova et al. 2001]{rhwt01} Ryutova M., Habbal S., Woo
R., \& Tarbell T. 2001, Sol. Phys., 200, 213

\bibitem[Saito et al. 2001]{sks01} Saito T., Kudoh T., \& Shibata K.
2001, ApJ, 554, 1151

\bibitem[Tagger \& Pellat 1999]{tp99} Tagger M., \& Pellat R. 1999,
A\&A, 349, 1003

\bibitem[Tagger et al. 1992]{tpc92} Tagger M., Pellat R., \& Coroniti
F. V. 1992, ApJ, 393, 708

\bibitem[Tatsuno et al. 2001]{tvy01} Tatsuno T., Volponi F.,
\& Yoshida Z. 2001,  Phys. Plasmas, 8, 399

\bibitem[Varni{\`{e}}r \& Tagger 2002]{vt02} Varni{\`{e}}r P., \& Tagger M. 2002,
A\&A, 394, 329

\bibitem[Velli \& Liewer 1999]{vl99} Velli M., \& Liewer P. 1999,
Space Sci. Rev., 87, 339

\bibitem[Volponi et al. 2000]{vyt00} Volponi F., Yoshida Z.,
\& Tatsuno T. 2000, Phys. Plasmas, 7, 2314

\bibitem[van der Holst et al. 2000]{vbb00} van der Holst B., Belien
A.J.C., \& Goedbloed J.P. 2000, Phys. Plasmas, 7, 4208

\end{thebibliography}
\end{document}